\documentclass[10pt,a4paper,final]{iopart}

\usepackage[utf8]{inputenc} 

\usepackage{iopams}  
\usepackage{graphicx}
\usepackage[breaklinks=true,colorlinks=true,linkcolor=blue,urlcolor=blue,citecolor=blue]{hyperref}

\usepackage{amsfonts}
\usepackage{gensymb}

\expandafter\let\csname equation*\endcsname\relax
\expandafter\let\csname endequation*\endcsname\relax
\usepackage{bm,amssymb}

\usepackage{braket}

\usepackage[numbers,sort&compress]{natbib}

\begin{document}

\title[Dynamical mean-field ab initio study of Vanadium diselenide monolayer ferromagnetism]{Dynamical mean-field study of Vanadium diselenide monolayer ferromagnetism}

\author{Taek Jung Kim$^{1}$, Siheon Ryee$^{1}$, Myung Joon Han$^{1*}$ and Sangkook Choi$^{2*}$ }

\address{$^{1}$Department of Physics, KAIST, 291 Daehak-ro, Yuseong-gu, Daejeon 34141, Republic of Korea}
\address{$^{2}$Condensed Matter Physics and Materials Science Department, Brookhaven National Laboratory, Upton, NY 11973, USA}
\ead{mj.han@kaist.ac.kr, sachoi@bnl.gov}

\vspace{10pt}
\begin{indented}
\item[]November 2019
\end{indented}

\begin{abstract}
To understand the magnetism of VSe$_2$, whose monolayer form has recently been reported to be a room temperature ferromagnet, Here, the combined method of conventional density functional theory with dynamical mean-field theory has been adopted. This higher-level computation method enables us to resolve many of existing controversies and contradictions in between theory and experiment. First of all, this new approach is shown to give the correct magnetic properties of both bulk and two-dimensional limit of VSe$_2$ which demonstrates its superiority to the conventional methods. The results demonstrate that monolayer VSe$_2$ without charge density waves is a ferromagnet with ordering temperature of 250K. From the direct simulation of temperature-dependent magnetic susceptibility and ordered moment, it is shown that its ferromagnetism is clearly two-dimensional in nature.  Further, it is shown that this ferromagnetic order is vulnerable to extra charge dopings which provides the important insight to elucidate recent experimental controversies. 
\end{abstract}

%
%
%
\maketitle
%
\ioptwocol

\section{Introduction}

Two-dimensional (2D) magnetism has played central roles in numerous scientific and technological advancements including high temperature superconductivity, quantum Hall related phenomena, and the development of new devices.\cite{lee_doping_2006,fernandes_what_2014-1,scalapino_common_2012,paglione_high-temperature_2010-2,nagaosa_anomalous_2010,stormer_fractional_1999,von_klitzing_quantized_1986,li_intrinsic_2019,liu_intrinsic_2018,cardoso_van_2018,huang_electrical_2018,jiang_controlling_2018,jiang_electric-field_2018,seyler_ligand-field_2018,wang_very_2018,wang_electric-field_2018,song_voltage_2019} In the past couple of years, notable magnetic orders have been observed even in single atomic layers. \cite{lee_ising-type_2016,huang_layer-dependent_2017,fei_two-dimensional_2018,bonilla_strong_2018,ohara_room_2018} These 2D magnetic van der Waals materials are expected to open up a wide range of possibilities. The vast phase space with different elements and structures suggests the straightforward tuning of its magnetic properties. Moreover, due to the ease of fabricating 2D heterostructures, magnetic van der Waals materials have a great potential to provide a new unit for multifunctional devices. \cite{zhong_van_2017,kim_one_2018,song_giant_2018,gong_two-dimensional_2019}

Among them, one of the fascinating materials is 1T-VSe$_2$.\cite{bayard_anomalous_1976,van_bruggen_magnetic_1976,strocov_three-dimensional_2012,barua_signatures_2017,pasztor_dimensional_2017,bonilla_strong_2018,chen_unique_2018,duvjir_emergence_2018,feng_electronic_2018} This layered  compound is composed of the ABC-stacked Se-V-Se atomic sheets (Figure~\ref{Figure_1}(a)). Since each V atom is surrounded by six Se, ionically V-$d$ states are split into $t_{2g}$ and $e_{g}$ manyfolds. Due to the trigonal distortion, $t_{2g}$ degeneracy is further lifted to form higher-lying $a_{1g}$ and two-fold degenerate {$ e_{g\pm}^{\pi}$}  states as depicted in Figure~\ref{Figure_1}(a). Most interestingly, VSe$_2$ monolayer is claimed to be a 2D ferromagnet with Curie temperature ($T_c$) above room temperature. \cite{bonilla_strong_2018,duvjir_emergence_2018,yu_chemically_2019} It is particularly interesting because bulk VSe$_2$ remains paramagnetic down to low temperature. \cite{bayard_anomalous_1976,van_bruggen_magnetic_1976,cao_defect_2017,barua_signatures_2017}

However, this conclusion remains elusive both theoretically and experimentally. For example, the reported experimental moments of $\approx15\mu_{B}$\cite{bonilla_strong_2018} or $5\mu_{B}$ \cite{duvjir_emergence_2018} are too large to be compared with the calculated values of about $0.6\mu_{B}$. \cite{ma_evidence_2012,li_versatile_2014,fumega_absence_2018} While a more recent experiment reports a much reduced moment of $\approx0.31\mu_{B}$ \cite{yu_chemically_2019}, the inconsistency needs to be further studied.
Furthermore, other experimental studies report the absence of ferromagnetic order for the same material, \cite{feng_electronic_2018,chen_unique_2018} which certainly requires a thorough theoretical investigation. The claim of monolayer ferromagnetism heavily relies on the first-principles calculation result based on DFT-GGA (density functional theory within generalized gradient approximation) which consistently produces the ferromagnetic ground state. \cite{ma_evidence_2012,li_versatile_2014} An obvious problem is, however, that DFT-GGA predicts the ferromagnetic phase also for the bulk VSe$_2$ being in a sharp contrast to the experimental fact of paramagnetism. \cite{lebegue_two-dimensional_2013,li_versatile_2014,bayard_anomalous_1976,van_bruggen_magnetic_1976,cao_defect_2017,barua_signatures_2017}

In order to perform a reliable investigation, one needs the higher-level theoretical framework suited for both itinerant and localized characters of electrons associated with the open-$d$ shells. First-principles methods in combination with dynamical mean-field theory (DMFT) provide an efficient way to study the dual nature of electrons by mapping a quantum many-body lattice problem onto a multi-orbital quantum impurity problem in an effective electron bath. \cite{georges_dynamical_1996-2} We hereby adopt the charge self-consistent LDA+DMFT to investigate the magnetic properties of VSe$_2$. For the construction of DMFT computations, the on-site Coulomb interaction parameters (Hubbard $U$ and Hund $J_H$) associated with V-$d$ orbitals have been calculated within constrained random phase approximation (cRPA) \cite{aryasetiawan_frequency-dependent_2004-1,sasioglu_effective_2011,choi_first-principles_2016}, and the nominal double-counting scheme \cite{PhysRevB.76.235101} is used for the self-energy correction (see Figure~\ref{Figure_1}(b, c) and Computational methodology section for more details).

\section{Computational methods}

\begin{figure}[h]
	\begin{center}
		\includegraphics[width=1.0\columnwidth]{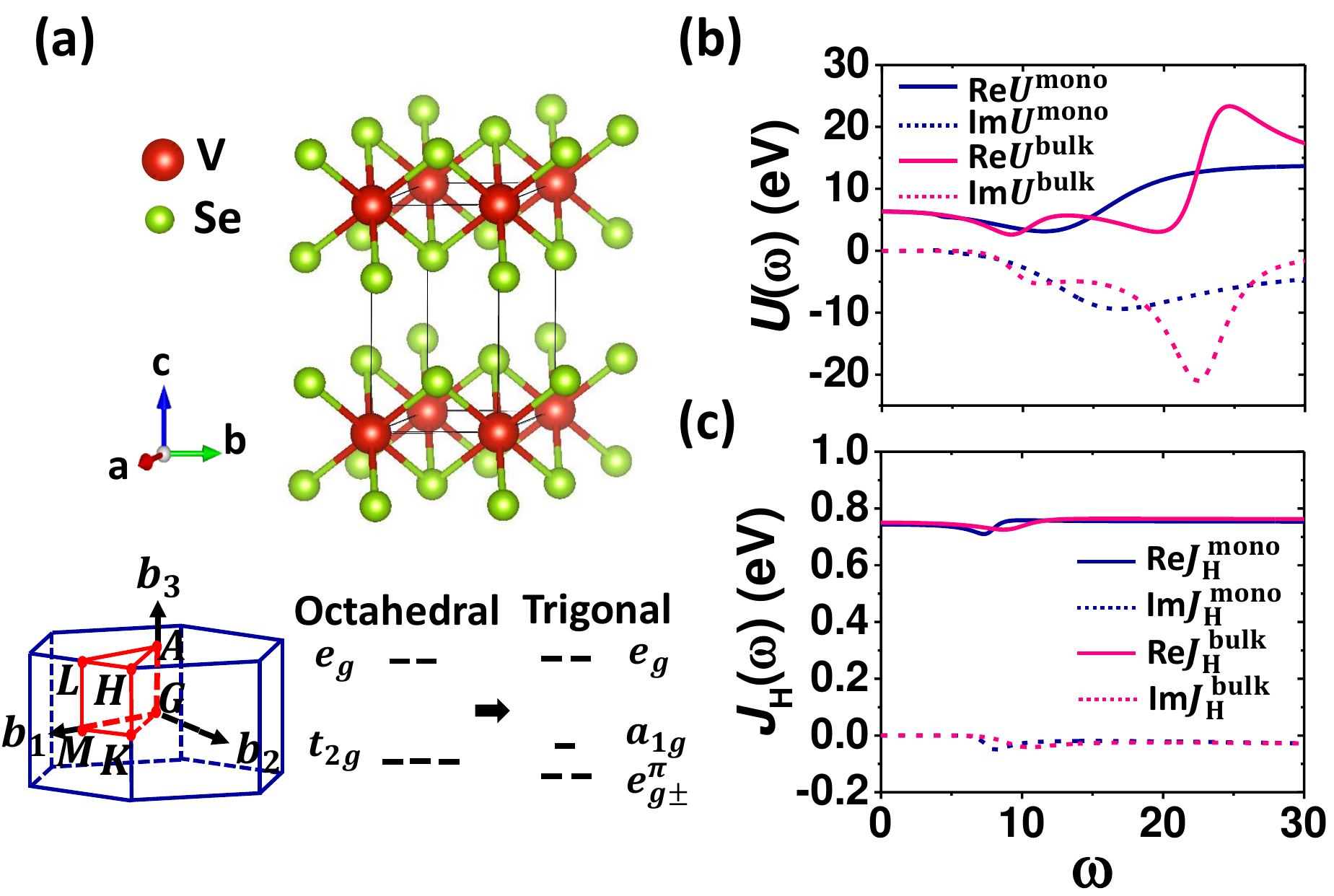}
		\caption{
			(a) Crystal structure of bulk 1T-VSe$_{2}$, high symmetry points in the first Brillouin zone, and
			schematic picture for the crystal field level splitting. Red and green spheres represent vanadium and selenium atom respectively. Due to the trigonally distorted octahedron, $t_{2g}$ states split into $a_{1g}$ and $e_{g\pm}^{\pi}$. (b, c) The $\omega$-dependent Coulomb (b) and Hund (c) interaction parameter calculated by cRPA for both monolayer (blue) and bulk (magenta) VSe$_{2}$. Solid and dashed lines present the real and imaginary parts, respectively.
			\label{Figure_1}}
	\end{center}
\end{figure}

The calculations have been performed by all-electrons DFT code Wien2K \cite{schwarz_augmented_nodate}, EDMFTF \cite{haule_dynamical_2010}, and ComDMFT package\cite{choi_comdmft_2018}. GGA functional parameterized by Perdew, Burke, and Ernzerhof \cite{perdew_generalized_1996} was adopted. The k-point grids of 16$\times$16$\times$7  and 26$\times$26$\times$1 were used for bulk and monolayer VSe$_{2}$, respectively. Crystal structure was determined by internal relaxation within GGA functional with the fixed experimental cell parameters \cite{bayard_anomalous_1976,van_bruggen_magnetic_1976,thompson_magnetic_1978,feng_electronic_2018,liu_epitaxially_2018,umemoto_pseudogap_2019} and the force criterion of 0.5 mRyd per \AA. For monolayer calculation, the vacuum layer of 30\AA~ has been taken into account. For on-site Coulomb interaction parameters, we used the cRPA technique \cite{aryasetiawan_frequency-dependent_2004-1,sasioglu_effective_2011,choi_first-principles_2016}  as implemented in ComDMFT package. \cite{choi_comdmft_2018} For the calculation of the partially screened Coulomb interaction, we chose the transitions between the bands in the hybridization energy window from $-$10 to +4 eV, and defined the polarizability from the transitions as our low-energy polarizability.

{Figure~\ref{Figure_1}(b, c) show the $\omega$-dependent Hubbard $U$ and Hund $J_H$ as obtained from our cRPA. As well known, $J_H$ is almost $\omega$-independent \cite{,PhysRevB.77.085122,PhysRevB.86.165105}. While $U(\omega)$ exhibits the more $\omega$ dependence, its behavior becomes nearly flat in the small $\omega$ regime which supports the conventional way of using static $U(\omega=0)$ value. Considering all these facts and following most of literature, we adopted the static value of $U$ and $J_H$ in the current study. We also checked the $U$ and $J_H$ dependence of our conclusion by varying $\pm$20\% of these interaction parameters from cRPA values. It is found that none of our conclusion is changed. The main change is the slight enhancement of moment formation when the larger values are used as expected. The change of (monolayer) moment at 100K is about 0.03$\mu_{B}$ and 0.25$\mu_{B}$ for the $U$ and $J_H$ change, respectively. It is also noted that the effect of $\omega$-dependence can effectively be understood by the use of slightly greater static $U$ value \cite{PhysRevB.91.125142,PhysRevLett.109.126408}.

The nominal double counting scheme has been adopted with 3 electron occupation in the V-d orbitals as predicted by DFT-GGA. \footnote{See also the experimental report by Cao {\it et al.} which shows that the valance electron number in VSe$_{2}$ is closer to the vanadium metal ($d^{3}$) rather than that of VO$_2$ ($d^{1}$)} \footnote{ We have also repeated calculations with another widely-used double counting choice, namely, `fully localized limit (FLL)' functional. While the use of FLL seems to slightly enhance ferromagnetism by producing a slightly larger monolayer moment of 0.438$\mu_{B}$ at 100K, no other noticeable change has been found.}  In order to properly describe the trigonally distorted local environment around V-d orbitals, we adopted a local basis set that diagonalizes the matrix $\epsilon_{imp}+\mathrm{Re}\triangle(\omega=0)$ (where, $\epsilon_{imp}:$ impurity levels, $\triangle(\omega)$ : hybridization function between local impurity and bath) during LDA+DMFT charge self-consistent calculation. Our LDA+DMFT calculations were performed by using Wien2K+EDMFTF. We also double-checked the result of paramagnetic phases using ComDMFT built on top of FlapwMBPT code \cite{KUTEPOV2017407} for \textit{ab initio} LDA calculations.

\section{Results and discussion}

\begin{figure*}[h]
	\begin{center}
		\includegraphics[width=2.0\columnwidth]{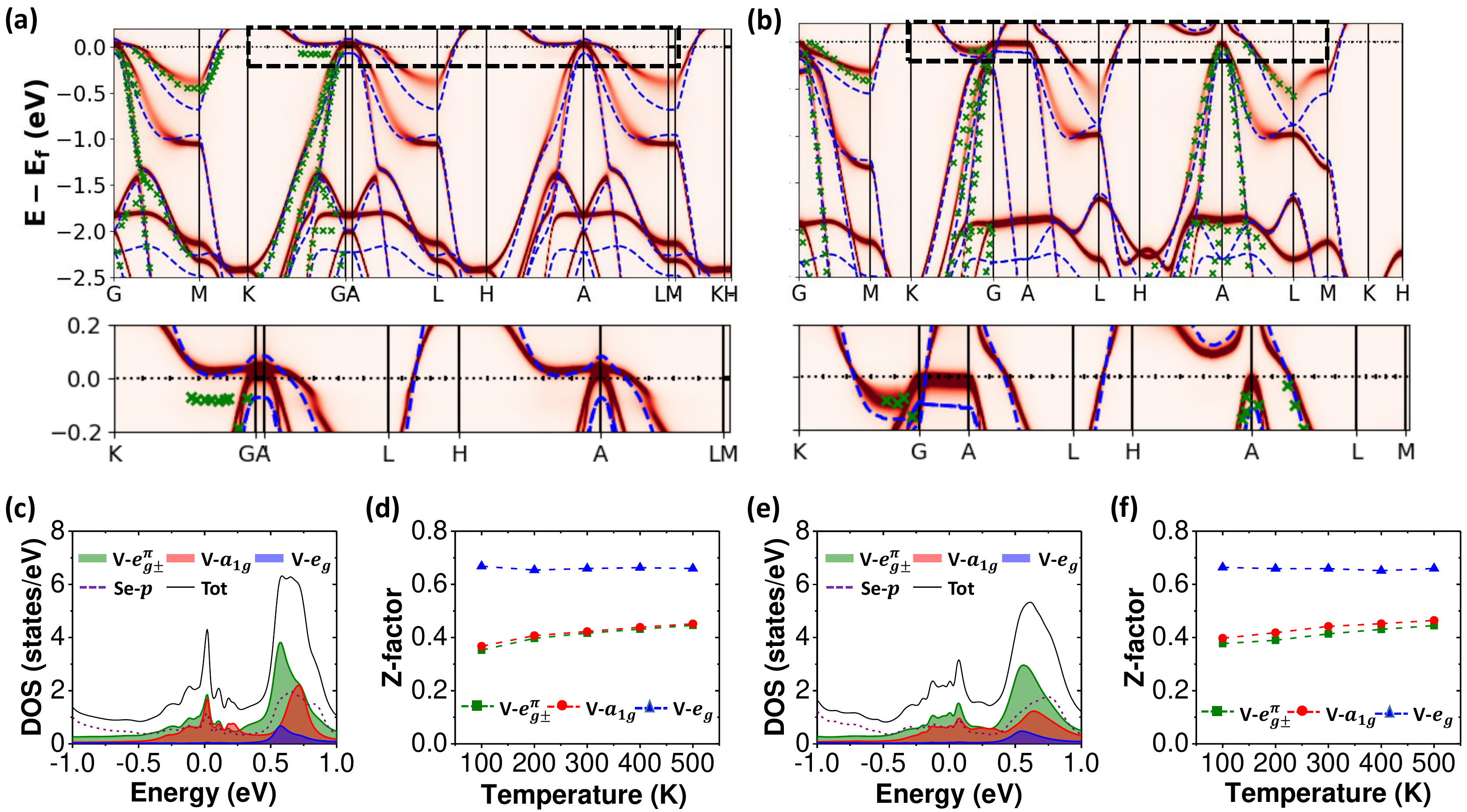}
		\caption{ The calculated electronic structure and the quasiparticle weight $Z$. (a, b) LDA+DMFT band dispersion of monolayer (a) and bulk (b)  VSe$_2$, respectively, in comparison with GGA and ARPES data. Red, blue and green colors represent the results of charge self-consistent LDA+DMFT spectral functions (at 300K), non-magnetic GGA, and ARPES \cite{feng_electronic_2018}. The black dashed box regions in the upper panels are enlarged in the lower panels. (c, e) The calculated PDOS of monolayer (c) and bulk (e) VSe$_2$ within charge-self-consistent LDA+DMFT. In the case of the monolayer, a well defined van Hove singularity peak appears at Fermi level whereas it is broaden in bulk. Green, red, blue colors represent $e_g^\pi$, $a$, and $e_g$ orbital states, respectively. The dashed and solids black lines show the Se-$p$ and total DOS, respectively. (d, f) The calculated quasiparticle weight associated with V-$d$ orbitals in the monolayer (d) and bulk (f) 1T-VSe$_2$. 
			\label{Figure_2}}
	\end{center}
\end{figure*}


Figure~\ref{Figure_2}(a, b) show the calculated electronic structure of monolayer and bulk VSe$_2$, respectively, in comparison with the previous angle-resolved photoemission spectroscopy (ARPES) data by Feng {\it et al.}~\cite{feng_electronic_2018}. Red, blue and green colors represent the result of LDA+DMFT in their paramagnetic phases, non-magnetic GGA, and ARPES spectra in the paramagnetic temperature regime, \cite{feng_electronic_2018} respectively. Two different theories coincidently predict that both monolayer and bulk 1T-VSe$_2$ are metals and that the interlayer coupling in bulk is not negligible; see, for example, the dispersive bands along G-A and K-H lines (Figure~\ref{Figure_2}(b)).

It should be noted that the dynamical local correlation (which is missing in LDA or GGA  calculation) plays an essential role for the description of important details around the Fermi level. Figure~\ref{Figure_2}(a, b) clearly show that LDA+DMFT results are in better agreement with experiments. See, for example, the G-M line in the range of $-$0.5-0.0 eV; the ARPES result (green crosses) of the highest valence band significantly deviates from the GGA band structure (blue dashed lines), exhibiting the larger (smaller) effective mass (band velocity). We also found that the calculation result of modified Becke-Johnson (MBJ) functional \cite{,becke_simple_2006-1,tran_accurate_2009} is similar with that of LDA/GGA. These notable features related to the electronic correlation are well-captured by DMFT (the red colored). The renormalized bands are mainly of V-$e_{g\pm}^\pi$ and V-$a_{1g}$ orbital character as demonstrated by projected density of states (PDOS) in Figure~\ref{Figure_2}(c, e). The effect of correlation can be quantified by the quasiparticle weight defined as $Z=(1-\frac{\partial{\textrm{Im}\Sigma_{ii}(i\omega_{n})}}{\partial{\omega_{n}}}|_{\omega_{n}=\frac{\pi}{\beta}} )^{-1}$ where $\Sigma_{ii}$ is the electronic self-energy associated with the orbital $i$ and $\beta$ is the inverse temperature; The band velocity and effective mass is renormalized by factor of $Z$ and $1/Z$, respectively. Figure~\ref{Figure_2}(d, f) show the calculated $Z$ factors for each V-$d$ orbital in 1T-VSe$_2$ monolayer and bulk, respectively. It is clearly shown that the correlation strengths in both systems are strongly orbital dependent. The $Z$ factors for two V-$e_g$ orbitals (blue triangles) are about 0.7 and basically temperature independent. In contrast, $Z$ for V-$e_{g\pm}^\pi$ (green squares) and V-$a_{1g}$  (red circles) are significantly smaller, $\sim$0.4, and decrease upon cooling in both bulk and monolayer.

\begin{figure}[h]
	\begin{center}
		\includegraphics[width=1.0\columnwidth]{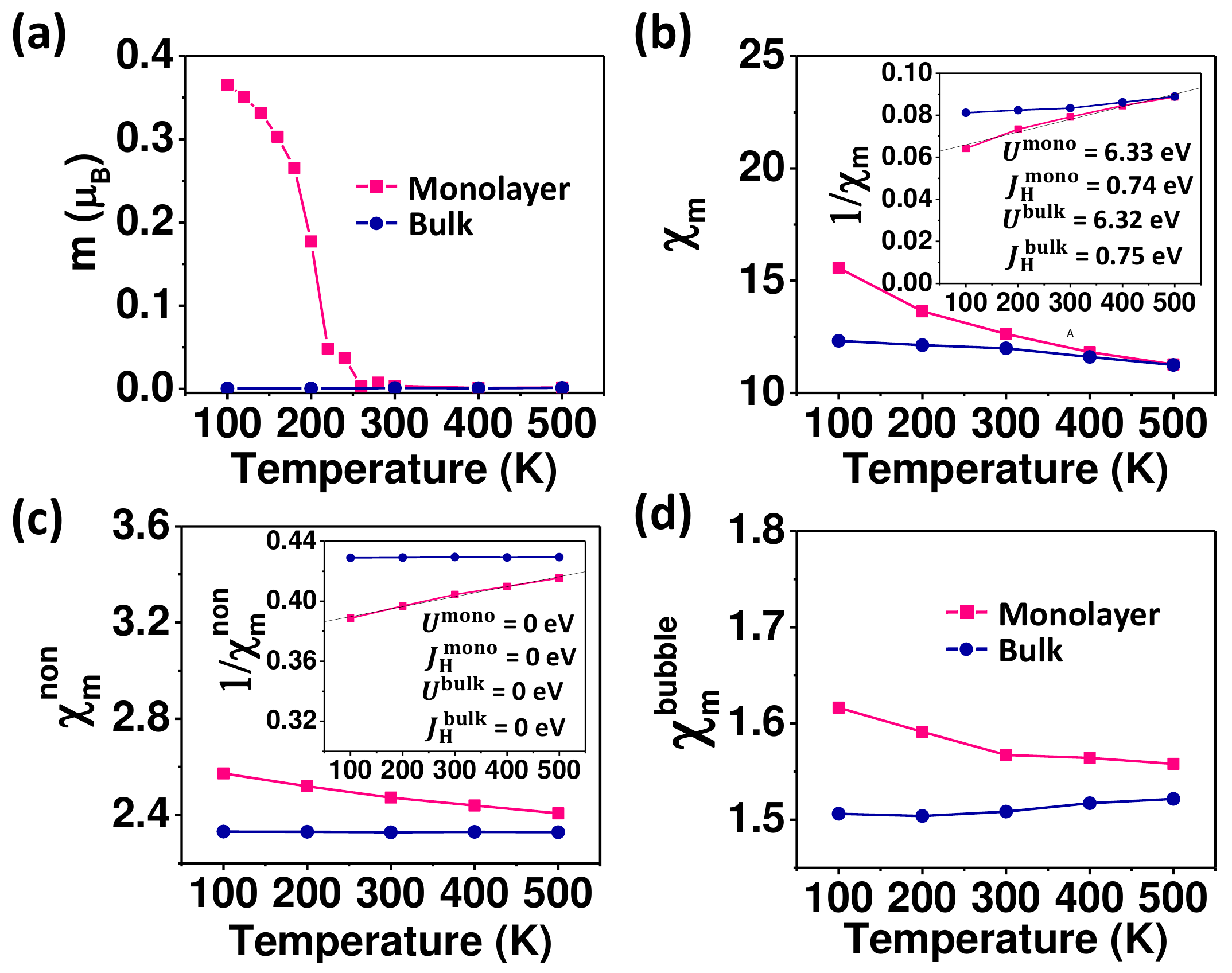}
		\caption{
			(a) The calculated ferromagnetic order parameter $m=\mu_B(\langle n_{\uparrow}\rangle -\langle n_{\downarrow} \rangle )$ as a function of temperature. The magenta line with squares and the blue with circles present the result for monolayer and bulk VSe$_{2}$, respectively. For monolayer, ferromagnetic transition is clearly identified at around 250K. (b) The calculated local spin susceptibility $\chi_{m}=\int_0^\beta d\tau \langle m(\tau)m(0)\rangle$ as a function of temperature. The inset shows the inverse susceptibilities which clearly show Curie- and Pauli-like behavior for monolayer and bulk, respectively. (c) The calculated non-interacting local spin susceptibility ($U=J_H=0$) of bulk (blue circles) and monolayer (magenta squares) in their paramagnetic phases. The inset presents its inverse. (d) The spin susceptibility calculated within bubble approximation.}
		\label{Figure_3}
	\end{center}
\end{figure}

To resolve the key issue in this material, namely, the possible monolayer ferromagnetism, we calculated the ferromagnetic order parameter $m=\mu_B(\langle n_{\uparrow}\rangle -\langle n_{\uparrow} \rangle )$ (where $\mu_B$  is Bohr magneton and $\langle n_{\uparrow(\downarrow)}\rangle$ is the up(down)-spin electron occupation in the V-$d$ subshell) as a function of temperature; see Figure~\ref{Figure_3}(a). The magenta squares and the blue triangles represent the LDA+DMFT results for monolayer and bulk VSe$_2$, respectively. The first thing to be noted is the absence of ferromagnetic order for the bulk phase as known from experiments. Note that GGA predicts the ferromagnetic order even for bulk, which significantly undermines its predictability for the magnetic ground state of monolayer VSe$_2$. \cite{lebegue_two-dimensional_2013,li_versatile_2014,ma_evidence_2012}

For monolayer, our LDA+DMFT calculation clearly shows that ferromagnetic order is developed and the magnetic moment is about 0.37$\mu_B$. In this regard, our result is seemingly in better agreement with the previous experimental reports of monolayer ferromagnetism \cite{bonilla_strong_2018,duvjir_emergence_2018,yu_chemically_2019}  rather than paramagnetism. \cite{feng_electronic_2018,chen_unique_2018} Also, it is in good agreement with a recent XMCD (x-ray magnetic circular dichroism) measurement \cite{yu_chemically_2019} rather than the older experimental reports \cite{bonilla_strong_2018,duvjir_emergence_2018}. As for the critical temperature, however, the calculated $T_c\approx250$K is markedly lower than the experimental value of $T_c \geq 300$K  \cite{bonilla_strong_2018,duvjir_emergence_2018,yu_chemically_2019}  These differences might reflect the presence of charge density wave (CDW) phase whose temperature range varies from 120K to 360K  depending on experiments. \cite{bonilla_strong_2018,feng_electronic_2018,duvjir_emergence_2018,umemoto_pseudogap_2019} There can be certain types of extrinsic effects in experimental situations such as defects and/or the interaction with substrate. Or, the difference of about 50K can be reconciled simply by taking slightly greater $U$ or/and $J_H$ value than cRPA ones. Considering the discrepancy even among the experimental $T_c$ and the various unexplored possibilities, any quantitative comparison needs to be careful. In the below we pursue the deeper understanding of magnetism in VSe$_2$.

Figure~\ref{Figure_3}(b) presents the temperature dependent local spin susceptibility $\chi_{m}=\int_0^\beta d\tau \langle m(\tau)m(0)\rangle$ (where $\tau$ is an imaginary time) which clearly shows the localized moments formed in the monolayer. Magenta squares and blue circles are the result of monolayer ($\chi_{m}^{\rm mono}$) and bulk ($\chi_{m}^{\rm bulk}$) susceptibility, respectively. For the latter, Pauli-like temperature-independent behavior demonstrates the electron delocalization in bulk VSe$_{2}$, being consistent with experimental report. \cite{cao_defect_2017} For the former, on the other hands, $\chi_{m}^{\rm mono}$ exhibits the weak but clear inverse-temperature dependence, indicative of the electron localization (see the inset). This local nature is another important new aspect that LDA+DMFT reveals for VSe$_2$ magnetism. We also found that the localized moment is formed mainly in the V-$e_{g\pm}^\pi$ and V-$a_{1g}$ orbitals.

In the following paragraphs, we argue that this local behavior in the monolayer VSe$_2$ is attributed to the concerted effect of reduced dimensionality and electronic correlation. Figure~\ref{Figure_3}(c) shows the calculated non-interacting spin susceptibilities ($\chi_{m}^{\rm non}$; $U$=$J_H$=0) in monolayer (magenta squares) and bulk (blue circles) in their paramagnetic phases. Note that the bulk susceptibility ($\chi_{m, {\rm bulk}}^{\rm non}$)  exhibits the Pauli-like behavior whereas the inverse-temperature dependence is clearly identified for the monolayer susceptibility ($\chi_{m, \rm {mono}}^{\rm non}$; see the inset). Note that this behavior of $\chi_{m, \rm{mono}}^{\rm non}$ is solely attributed to the band structure (without interaction effect), particularly to the van Hove singularity near the Fermi level. \cite{mielke_ferromagnetism_1993,ulmke_ferromagnetism_1998,wahle_microscopic_1998,han_ferromagnetism_2016,hausoel_local_2017} As shown in Figure~\ref{Figure_2}(a), van Hove singularity is strongly enhanced in the monolayer VSe$_2$, implying the vanishing electron velocity; compare the dispersions along K-G and H-A line in Figure~\ref{Figure_2}(a) to \ref{Figure_2}(b). This electronic behavior naturally leads to the saddle-point formation \footnote{The van Hove singularity of monolayer VSe$_{2}$ is in the G--K and A--H line along which the near-Fermi-level band  makes a minimum while it does a maximum along the orthogonal direction to those lines, forming a saddle point} and the divergence in DOS. For the bulk, on the other hand, the inter-layer coupling is noticeable; for example, compare the dispersion along the M-K-G path to that along H-A-L (Figure~\ref{Figure_2}(a, b)). It naturally weakens the van Hove singularity leading to the broader DOS near the Fermi energy as shown in Figure~\ref{Figure_2}(c, e). This characteristic band feature of monolayer VSe$_2$ is the effect of reduced dimensionality.

The local dynamical correlation plays a crucial role on top of the pre-localization of electrons caused by two dimensionality. We found that the correlation enhances the electron localization in two distinctive ways. First, the local dynamical correlation renormalizes the electronic structure, especially the bands with the critical point. The orbital character of these bands are mainly V-$e_{g\pm}^\pi$ and V-$a_{1g}$  as shown in Figure~\ref{Figure_2}(d, f). Their $Z^{-1}$ values greater than two demonstrate the strong enhancement of electron localization caused by the band structure renormalization. It can further be confirmed by comparing $\chi_{m}^{\rm non}$ to the spin susceptibility within bubble approximation; $\chi_{m}^{\rm bubble}=2\mu_B^2\sum_{i,j\in{V-}d}\int_0^\beta d\tau G_{i,j}(\tau)G_{j,i}(-\tau)$ (where $G_{i,j}$ is a LDA-DMFT local Green's function). As shown in Figure~\ref{Figure_3}(c, d), $\chi_{m}^{\rm non}$ is almost two times greater than $\chi_{m}^{\rm bubble}$. It is consistent with the $Z^{-1}$ value of V-$e_{g\pm}^\pi$ and V-$a_{1g}$ orbitals, demonstrating that this enhancement originates from the band renormalization. Second, in addition to the quasiparticle band structure effect, the dynamical correlation facilitates the electron localization in the two-particle level. Compared to $\chi_{m}^{\rm bubble}$, the spin susceptibility, $\chi_{m}$, is strongly enhanced; compare the magenta line in Figure~\ref{Figure_3}(b) with that in Figure~\ref{Figure_3}(d). Its temperature dependence is also more pronounced than $\chi_{m}^{\rm bubble}$, indicative of the important contribution coming from  vertex correction to the spin susceptibility of vanadium.  

Here we stress that the reduced dimensionality is essential for the local moment formation as demonstrated by the fact that the bulk susceptibility, $\chi_{m,\rm{bulk}}$, remains temperature-independent; see Figure~\ref{Figure_3}(b). Namely, the electronic correlation alone cannot induce the moment formation even if the $\chi_{m, \rm{bulk}}$ is significantly greater $\chi_{m, \rm{bulk}}^{\rm non}$ in the entire temperature range. As the origin of local moment formation, the combined effect of dimensional reduction and electron correlation is not just physically interesting but also possibly important to understand experiments. For example, it implies that the local moment can be suppressed by the inter-layer hoppings or the monolayer-substrate interactions.

\begin{figure}[h]
	\begin{center}
		\includegraphics[width=1.0\columnwidth]{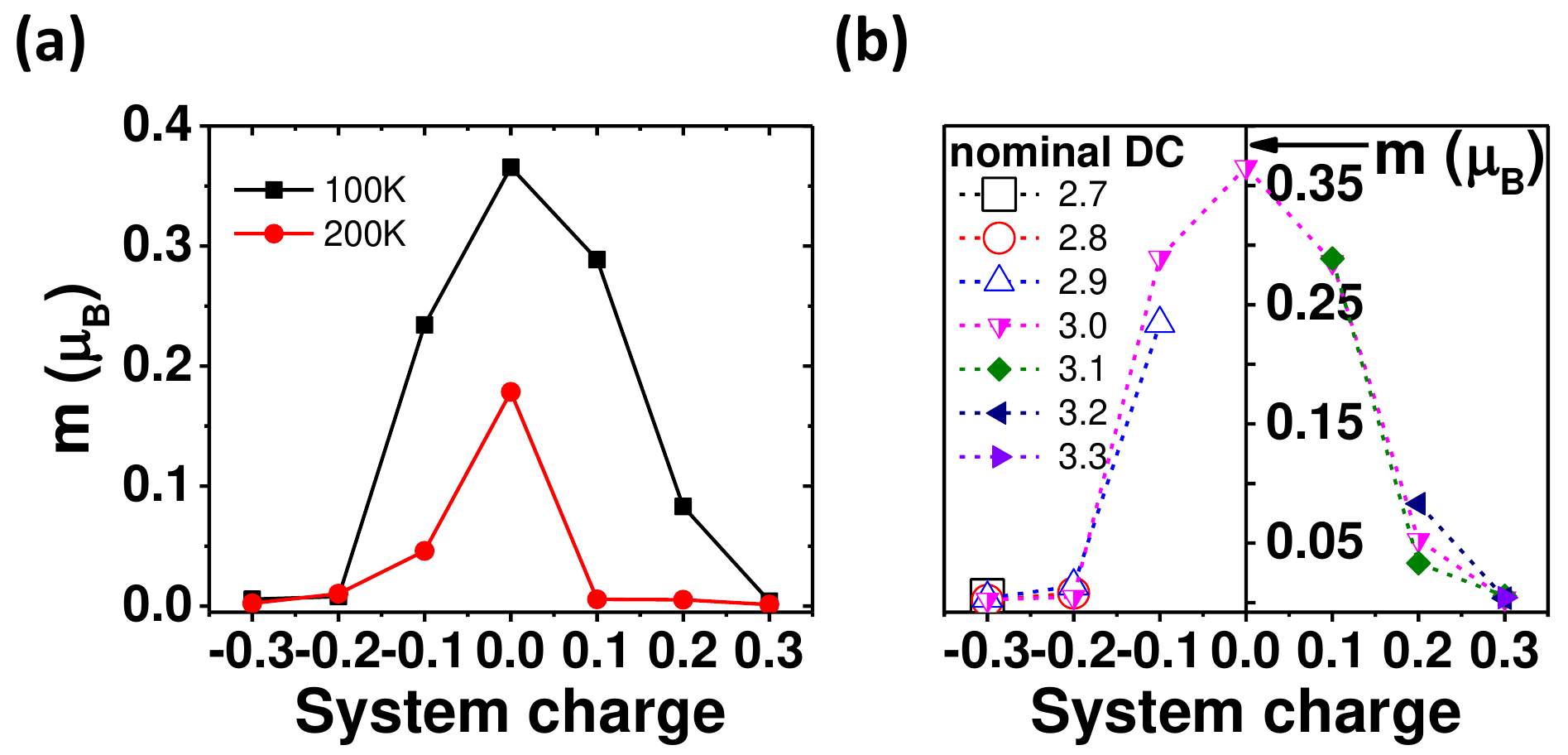}
		\caption{ (a) The calculated ferromagnetic order parameter $m=\mu_B(\langle n_{\uparrow}\rangle -\langle n_{\downarrow} \rangle )$ for monolayer VSe$_{2}$ with a varying system charge at two different temperatures. The hole-like and electron-like extra charges have been introduced through the DFT system charge option  (and the updated DMFT chemical potential accordingly). (b) The calculated $m$ with a varying system charge together with nominal double-counting charge at T=100K. The seven different values for V-$d$ occupations in the nominal double-counting choice have been considered ranging from $-$0.3 to +0.3 (the same with system charge range). 
			\label{Figure_4}}
	\end{center}
\end{figure}

In order to explore other possible reasons for the contradictory experimental reports on the monolayer ferromagnetism, \cite{chen_unique_2018,feng_electronic_2018,bonilla_strong_2018,duvjir_emergence_2018,yu_chemically_2019} one can consider both cation(or anion) defects and charge transfer from substrate as often observed in this type of chalcogenide systems.\cite{nguyen_atomic_2017,jang_origin_2019-1,mcqueen_extreme_2009,li_first-order_2009-1,han_doping_2009-1,barua_signatures_2017,feng_electronic_2018,verchenko_ferromagnetic_2015,may_magnetic_2016,liu_critical_2017,liu_anomalous_2018} In fact, our calculation shows that the ferromagnetic order is vulnerable to the extra effective charges introduced. Figure~\ref{Figure_4} presents the calculated ferromagnetic moment $m$ as a function of varying system charges. In Figure~\ref{Figure_4}(a), we simulate the effect of both hole-like (negative values) and electron-like (positive values) extra charges by tuning the system charge at the DFT level while, in Figure~\ref{Figure_4}(b), both DFT system charge and DMFT nominal double-counting value are simultaneously controlled.
The results demonstrate that introducing the extra charges through, for example, V or Se defects, can easily suppress the ferromagnetic order. The number of extra charges  ({\it i.e.}, $-0.3 \leq e \leq +0.3$) seems to be reasonable  to simulate the possible defect concentrations considering the reported values of related systems such as Fe$_3$GeTe$_2$ and FeTe. \cite{verchenko_ferromagnetic_2015,may_magnetic_2016,liu_critical_2017,liu_anomalous_2018,li_first-order_2009-1} This result hopefully provides useful information to understand the previous experimental reports and to stimulate the further investigations.

Finally, an important factor that has not been directly addressed in the current study is the cooperation or competition of ferromagnetism with CDW order. It is known that the monolayer VSe$_2$ also hosts the CDW order while the CDW vector can be different from its bulk counter part \cite{bonilla_strong_2018,feng_electronic_2018,duvjir_emergence_2018,chen_unique_2018,umemoto_pseudogap_2019}. Importantly, all experiments that observed ferromagnetic signal coincidentally report the  $T_c$ well above $T_{\rm CDW}$ \cite{bonilla_strong_2018,duvjir_emergence_2018,yu_chemically_2019}. Thus, our result of ferromagnetic order in the undistorted 1T structure is quite meaningful. It is noted simultaneously that the formation of CDW phase likely suppresses the ferromagnetic order rather than enhances as noted in a recent study \cite{fumega_absence_2018}.
Considering many different reports on the temperature scale as well as the pattern of CDW \cite{bonilla_strong_2018,feng_electronic_2018,duvjir_emergence_2018,umemoto_pseudogap_2019,akgenc_phase-dependent_2020}, this issue certainly requires further careful investigations in both theory and experiment.

\section{Conclusion}

We investigated the magnetic properties of bulk and monolayer VSe$_2$ by using charge self-consistent LDA+DMFT approach. Our results show that VSe$_2$ monolayer is a ferromagnetically-ordered material below 250K whereas its bulk phase is paramagnetic. The electron localization and subsequent local moment formation in the monolayer originate from the quasiparticle pre-localization enabled by the reduced dimensionality and its enhancement by local dynamical correlation. This study provides an important example where the reduced dimensionality is an essential factor to form local moments and subsequent magnetic orderings. Our work will be useful in the understanding and design of possible magnetic devices based on 2D heterostructures.

\section{Acknowledgements}
We thank J.-H. Sim and H.-S.Kim for useful technical discussion. T. J. Kim, S. Ryee and M. J. Han were supported by BK21plus program, Basic Science Research Program (2018R1A2B2005204) and Creative Materials Discovery Program through NRF (2018M3D1A1058754). T. J. Kim, S. Ryee and S. Choi were supported by the U.S Department of Energy, Office of Science, Basic Energy Sciences as a part of the Computational Materials Science Program. This research used resources of the National Energy Research Scientific Computing Center (NERSC), a U.S. Department of Energy Office of Science User Facility operated under Contract No. DE-AC02-05CH11231.

\bibliography{2dmat}
\bibliographystyle{iopart-num}

\end{document}